\documentclass[showpacs,preprintnumbers,superscriptaddress]{revtex4}
\usepackage{graphicx}
\usepackage{bm}

\begin{document}

\title{Charged-particle rapidity density in Au+Au collisions
in a quark combination model}

\author{Feng-lan Shao}
%\email{shaofl@mail.sdu.edu.cn}
\affiliation{Department of Physics, Qufu Normal University,
Shandong 273165, People's Republic of China}

\author{Tao Yao}
%\email{yao@sdu.edu.cn}
\affiliation{Department of Physics, Shandong University,
Shandong 250100, People's Republic of China}

\author{Qu-bing Xie}
%\email{xie@sdu.edu.cn}
\affiliation{Department of Physics, Shandong University,
Shandong 250100, People's Republic of China}

\begin{abstract}
Rapidity/seudorapidity densities for charged particles and their
centrality, rapidity and energy dependence in Au+Au collisions at
RHIC are studied in a quark combination model. Using a Gaussian-type
rapidity distribution for constituent quarks as a result of Landau
hydrodynamic evolution, the data at $\sqrt{s_{NN}}=130, 200$ GeV at
various centralities in full pseudorapidity range are well
described, and the charged particle multiplicity are reproduced as
functions of the number of participants. The energy dependence of
the shape of the $dN_{ch}/d\eta$ distribution is also described at
various collision energies $\sqrt{s_{NN}}=200, 130, 62.4$ GeV in
central collisions with same value of parameters except $19.6$ GeV.
The calculated rapidity distributions and yields for the charged
pions and kaons in central Au+Au collisions at $\sqrt{s_{NN}}=200$
GeV are compared with experimental data of the BRAHMS Collaboration.

\end{abstract}


\maketitle

\section{Introduction}

The relativistic heavy ion collider (RHIC) at Brookhaven National
Lab was built to search for quark matter, or the so-called
quark-gluon plasma (QGP). Since its first run in 2000, a huge number
of data have been accumulated and a comprehensive analysis of these
data has been carried out. A variety of experimental facts from
different aspects imply that the strongly coupled QGP has probably
been produced in central Au+Au collisions at RHIC. For recent
reviews of QGP and summary of experimental data, see e.g.
\cite{Adams:2005dq,Gyulassy:2004zy,Jacobs:2004qv,Kolb:2003dz,
Braun-Munzinger:2003zd,Rischke:2003mt}. Central Au+Au collisions are
characterized by the production of thousands of charged-particles in
vacuum. The charged-particle density per unit rapidity or
pseudorapidity $dN_{ch}/dy$ or $dN_{ch}/d\eta$ is one of the most
important observables to measure for the signal of QGP, from which a
lot of information about the hot and dense matter can be extracted
\cite{Wang:2000bf,Kharzeev:2000ph,Kharzeev:2001gp,Capella:2000dn,
DiasdeDeus:2000gf,Eskola:1999fc,Eskola:2002qz}. One can scale
$dN_{ch}/d\eta$ or $dN_{ch}/dy$ by the number of participant nucleon
pairs $\langle N_{\rm part}/2\rangle$ and observe its logarithmic
increase with $\langle N_{\rm part}\rangle$, which is regarded as an
evidence of color glass condensate
\cite{McLerran:1993ni,Venugopalan:2005mg,Kharzeev:2000ph,Kharzeev:2001gp}.
From the rapidity/pseudorapidity density and the transverse energy
per particle, one can determine via Bjorken method the real density
of the fireball which can provide one piece of evidence for the
deconfinement phase transition. The experimental data about the
charged-particle rapidity density have been presented by the PHOBOS
collaboration \cite{Back:2000gw,Back:2002wb}, the PHENIX
collaboration \cite{Adcox:2000sp}, and the BRAHMS Collaboration
\cite{Bearden:2001xw,Bearden:2001qq}.

In this paper we will use a quark combination model to study the
rapidity/pseudorapidity density varied with the number of
participants and the energy in full rapidity range. The quark
combination picture is successful in describing many features of
multi-particle production in hadronic collisions. In
ultra-relativistic heavy ion collisions at RHIC energies, a lot of
new features are found, e.g. the high ratio of $n_{p}/n_{\pi}\sim 1$
at intermediate transverse momenta, which supports quark coalescence
or recombination picture
\cite{Fries:2003vb,Greco:2003xt,Hwa:2002tu}. The quark number
scaling of the elliptic flow is also a manifestation of the quark
coalescence or recombination
\cite{Voloshin:2002wa,Molnar:2003ff,Lin:2002rw}. In this paper we
will use a binary potential model for the constituent quark
production and then let the constituent quarks combine into initial
hadrons according to a quark combination rule. Then we allow the
resonances in the initial hadrons to further decay to final hadrons
with the help of the event generator  PYTHIA 6.3 \cite{Sjostrand}.

The paper is organized as follows. In the next section we give a
brief description of the model for constituent quark production and
combination. In section III, we present our predictions for the
rapidity/pseudorapidity densities varied with the number of
participants in the full rapidity range at $\sqrt{s_{NN}}=130, 200$
GeV, the energy dependence of the $dN_{ch}/d\eta$ distribution at
various collision energies for central collisions, and the results
for the rapidity densities $dN/dy$ and yields for charged pions and
kaons in the central collisions at $\sqrt{s_{NN}}=200$ GeV. The
summary and discussions are in section IV.

\section{The quark production and combination model}

In this section we give a brief introduction of
the quark production and combination model we use.
The model was first proposed for high energy $e^+e^-$ and
$pp$ collisions \cite{Xie:1988wi,Liang:1991ya,Wang:1995ch,
Zhao:1995hq,Wang:1996jy,Si:1997ux} and recently extended
to ultra-relativistic heavy ion collisions \cite{Shao:2004cn,Yao:2006fk}.
It has also been applied to
the multi-parton systems in high energy $e^+e^-$ annihilations
\cite{Wang:1995gx,Wang:1996pg,Wang:1999xz,Wang:2000bv}.

\subsection{An effective model for quark production}

The quark production from vacuum is a very sophisticated
non-perturbative process. The color glass condensate model is a
semi-classical QCD effective theory for the quark production in
heavy ion collisions \cite{McLerran:1993ni,Venugopalan:2005mg}. In
this paper we use a simple model for quark production which is of
statistical nature without dynamic details. We determine the number
of constituent quarks by the total effective energy for producing
quarks from the vacuum excitation. The effective energy consists of
the part for quark static masses and that for effective interquark
potentials.

Consider a system of $N_q$ quarks and anti-quarks excited in vacuum
, the number of light and strange quarks/anti-quarks follow the
ratio $N_u:N_d:N_s=1:1:\lambda _s$ with $N_q=N_u+N_d+N_s$, where
$\lambda _s<1$ is the strangeness suppression factor due to the
heavier mass of strange quarks/anti-quarks. The average quark mass
is given by $m=(2m_u+\lambda _s m_s)/(2+\lambda _s)$, where
$m_u=m_d$ is the light quark mass and $m_s$ the strange quark mass.

We assume that the interaction is characterized by an inter-quark
potential $V$ which takes a substantial fraction of total effective
energy. The constituent quark number can be determined from the
following energy equation,
\begin{equation}
\label{eq6}
E=\langle{N_q}\rangle m +\frac{\langle{N_q}\rangle}{2}
(\langle{N_q}\rangle-1)\langle{V}\rangle ,
\end{equation}
which gives the number of constituent quarks as
\begin{equation}
\label{eq7}
\langle{N_q}\rangle=2[(\alpha^{2}+\beta E)^{1/2}-\alpha],
\end{equation}
where
\begin{equation}
\label{eq8}
\beta\equiv\frac{1}{2\langle V\rangle}, \alpha\equiv\beta m-\frac{1}{4}.
\end{equation}
Note that the quark number $N_q$ follows a specific distribution, so
does the potential, we have taken their averages in the above
equations. In Eq. (\ref{eq6}), we only included the two-body
potential leading to a $E^{1/2}\sim s^{1/4}$ asymptotic behavior for
$N_q$ at high energy if $\langle V\rangle$ is constant. For a strong
coupling system, it is possible that the $n$-body ($n>2$) potential
might be more important, and the asymptotic behavior then becomes
$N_q\sim s^{1/2n}$. When $n$ is large, $N_q$ more and more
approaches a logarithmic increase with energy.

The PHOBOS experiments have shown that above SPS energies, the total
multiplicity per participant pair $\langle{N_{\rm
ch}}\rangle/\langle{N_{\rm part}/2}\rangle$ in central events scales
with $\sqrt{s_{NN}}$ in the same way as $e^+e^-$ collisions
\cite{Back:2002wb,Back:2003xk}. This suggests a universal mechanism
of particle production  mainly controlled by the amount of effective
energy available for particle production. Based on this property we
extend the above quark production model originally applied to
$e^+e^-$ annihilation and $pp$ collisions to heavy ion collisions.
The average quark number in nucleus-nucleus collisions can be
written as
\begin{equation}
\label{eq9}
\langle{N_q}\rangle=2[(\alpha^{2}+\beta\sqrt{s_{NN}})^{1/2}-\alpha]
\langle{N_{\rm part}}/2\rangle,
\end{equation}
where we have taken $E=\sqrt{s_{NN}}$. Note that the effective
energy $E$ is equal to collision energy $\sqrt{s}$  for light quark
events in $e^+e^-$ annihilation ( but $E\neq \sqrt{s}$ for heavy
quark events). Here we have assumed that the colliding nuclei is
fully stopped and all the colliding energy $\sqrt{s_{NN}}$ is used
for particle production, which is reasonable for the central
collisions due to the fully rescatterings of partons towards the
local equilibrium. It is also consistent with the Landau's space
time evolution picture \cite{Landau:1953gs}. Note that the average
number of quarks and antiquarks $\langle{N_q}\rangle$ includes not
only new produced quarks and antiquarks, but also some additional
quarks left by the incident nuclei.

\subsection{Model for quark combination}

In this subsection we briefly summarize how quarks combine into
hadrons in our model. Different from many coalescence models which
do not distinguish directly produced hadrons from final state
hadrons, our quark combination model only describe the hadronization
of initially produced hadrons including resonances. Then all
resonances are allowed to decay into final state hadrons. Here we
make use of the event generator PYTHIA 6.3 \cite{Sjostrand} to deal
with resonance decays. The basic idea is to put $N_q$ quarks and
anti-quarks line up in a one-dimensional order in phase space, e.g.
in rapidity, and let them combine into initial hadrons one by one
following a combination rule. See section II of Ref.
\cite{Shao:2004cn} for short description of such a rule. We note
that it is very straightforward to define the combination in one
dimensional phase space, but it is highly complicated to do it in
two or three dimensional phase space \cite{Hofmann:1999jx}. The
flavor SU(3) symmetry with strangeness suppression in the yields of
initially produced hadrons is fulfilled in the model
\cite{Xie:1988wi,Wang:1995ch}. Using the model, we have described
most of multiplicity data for hadrons in electron-positron and
proton-proton/anti-proton collisions
\cite{Xie:1988wi,Liang:1991ya,Wang:1995ch,Zhao:1995hq,Wang:1996jy,Si:1997ux}.
Also we solved a difficulty facing other quark combination models in
describing the TASSO data for baryon-antibaryon correlation in
electron-positron collisions \cite{Si:1997ux}. Combined with the
color flow picture \cite{Wang:1996pg}, the model can describe the
hadroniztion of multiparton states
\cite{Wang:1995gx,Wang:1999xz,Wang:2000bv}. We have extended the
model to reproduce the recent RHIC data for hadron multiplicity
ratios, $p_{T}$ spectra \cite{Shao:2004cn} and elliptic flows
\cite{Yao:2006fk} in central rapidity region.

%\begingroup
%\squeezetable
\begin{table}
\label{yield200GeV} \caption{Yields of charged pions and kaons
compared with BRAHMS data at 200 GeV \cite{Bearden:2004yx}.}
\tabcolsep0.15in \arrayrulewidth1pt
\begin{tabular}{|c|c|c|}
\hline
 & DATA & our model \\
\hline
$\pi^{+}$ & $1660\pm{15}\pm{133}$ & 1676\\
\hline
$\pi^{-}$ & $1683\pm{16}\pm{135}$ & 1680\\
\hline
$K^{+}$ & $286\pm{5}\pm{23}$ & 280\\
\hline
$K^{-}$ & $242\pm{4}\pm{19}$ & 242\\
%\hline
%$p$ & $138\pm{7}$ & 153\\
%\hline
%$\bar{p}$ & $84\pm{6}$ & 85\\
\hline
\end{tabular}
\end{table}
%\endgroup

\section{Rapidity and pseudorapidity densities}
\label{hadron}

In this section, we use our combination model to compute the
centrality dependence of distributions of rapidity/pseudorapidity
densities in Au+Au collisions at $\sqrt{s_{NN}}=130, 200$ GeV and
study the energy dependence of the shape of the $dN_{ch}/d\eta$
distribution at various collision energies $\sqrt{s_{NN}}=19.6,
62.4, 130$ and $200$ GeV for central collisions, and also calculate
the rapidity densities $dN/dy$ and yields for charged pions and
kaons in the central collisions at $\sqrt{s_{NN}}=200$ GeV.

First we have to fix the parameters of the model. There are two
parameters $m$ and $\langle V \rangle$ or $\alpha$ and $\beta$ in
Eq.(\ref{eq7}). As we pointed out that these quarks and antiquarks
are constituent ones, so we use constituent masses
$m_{u}=m_{d}=0.34$ GeV and $m_{s}=0.5$ GeV giving the average mass
$m=0.36$ GeV. The strangeness suppression factor is chosen to be
$\lambda_{s}=0.55$ by fitting the data at RHIC energies
\cite{Shao:2004cn}. The parameter $\beta$ is set to 3.6 GeV$^{-1}$
which described the $e^{+}e^{-}$ data. The parameters controlling
the total multiplicity are the number of quarks and that of
anti-quarks. In electron-positron and proton-antiproton collisions,
the number of quarks is equal to that of anti-quarks, i.e. there are
no excess baryons in contrast to anti-baryons. For nucleus-nucleus
collisions, however, there are some excess baryons deposited by the
colliding nuclei. The total number of quarks and anti-quarks
$\langle N_q \rangle$ is given by Eq.(\ref{eq9}). The number of net
quarks can be further determined by the ratio of anti-proton to
proton \cite{Shao:2004cn}. At 130 and 200 GeV, we find that the net
quark numbers are about 420 and 360, respectively.

\begin{figure}
\includegraphics[scale=0.4]{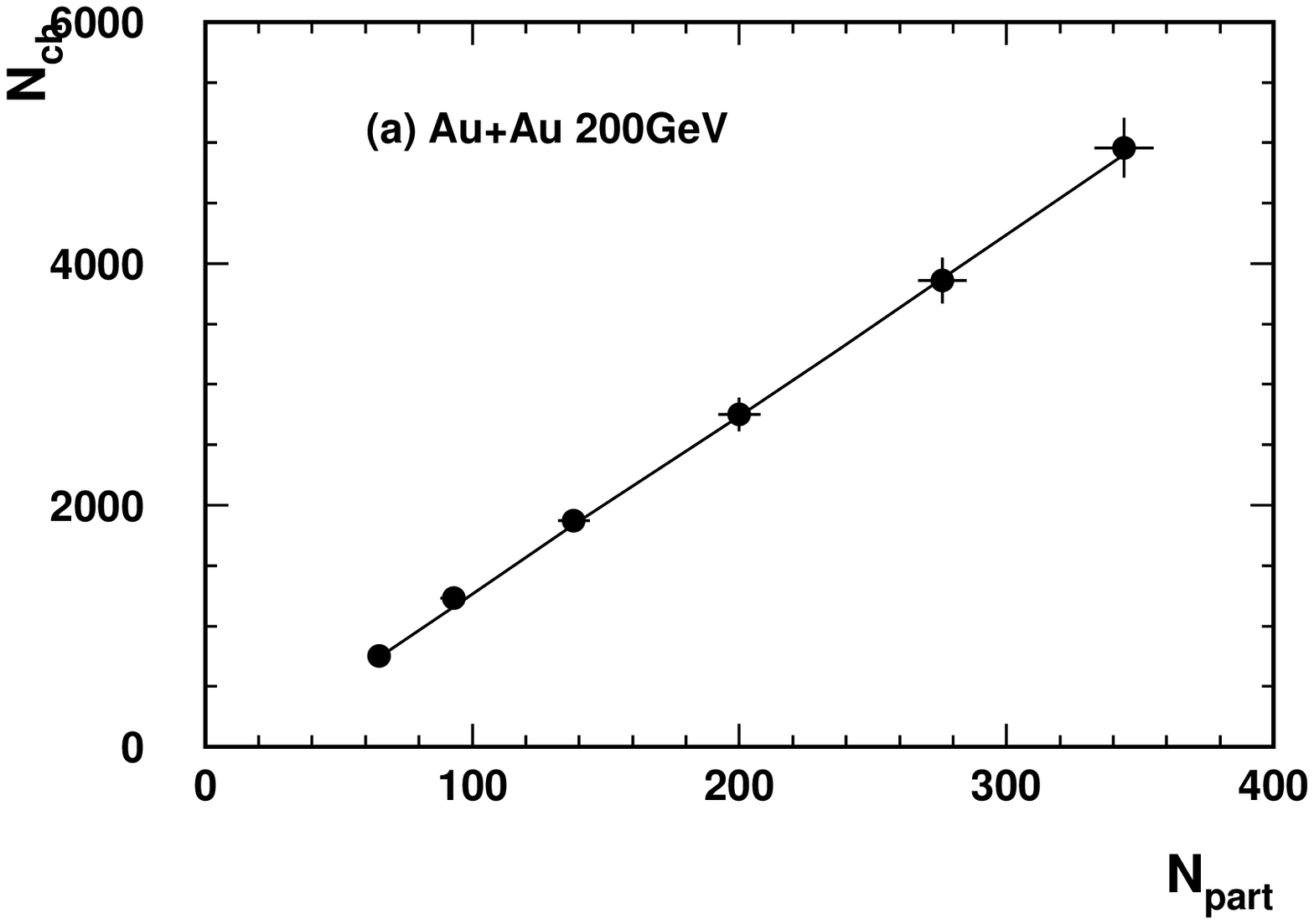}
\includegraphics[scale=0.4]{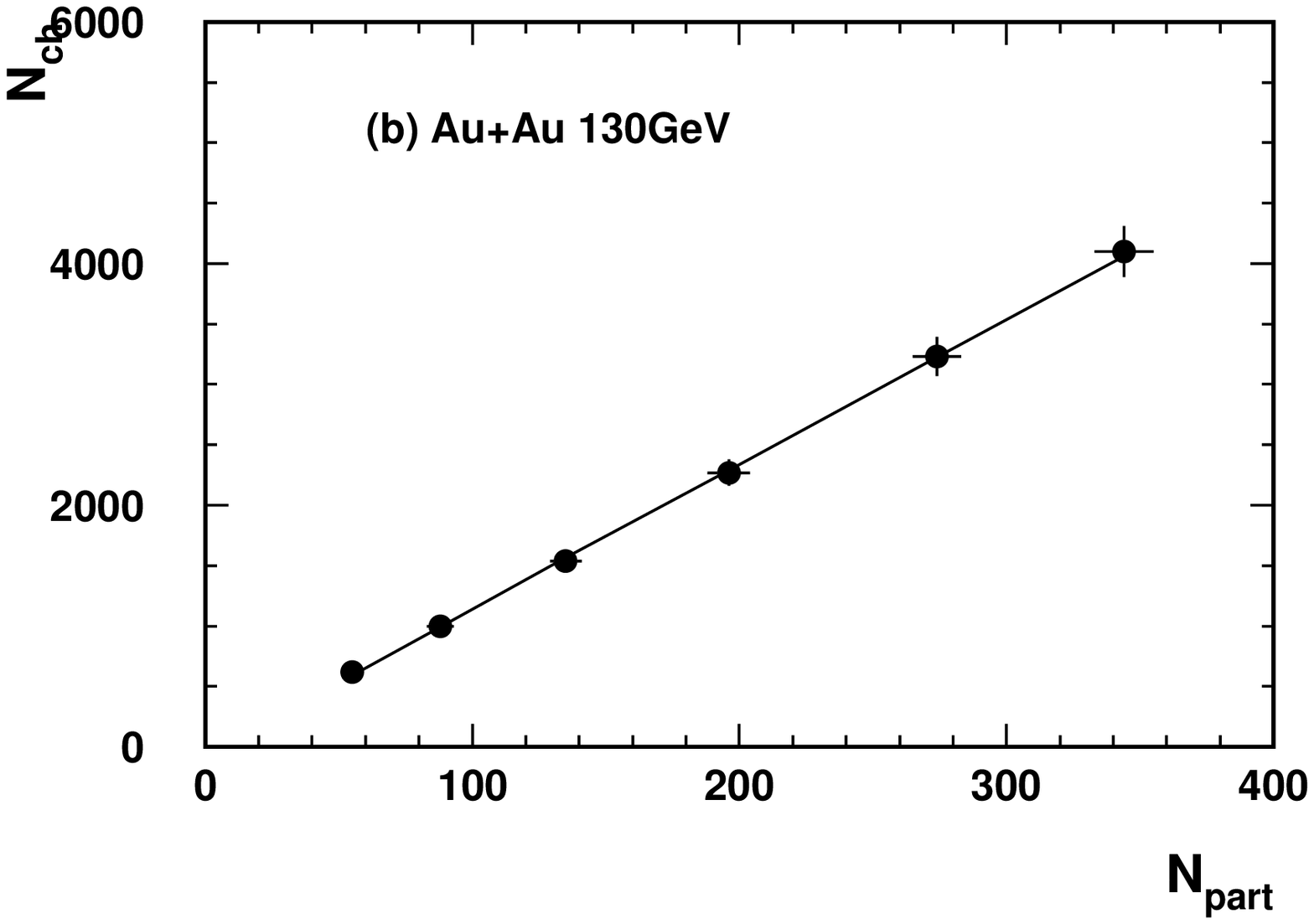}
\caption{The centrality-dependent multiplicities of charged
particles in Au+Au collisions at $\sqrt{s_{NN}}=130, 200$ GeV.
The solid lines are our results. The data are taken from PHOBOS
\cite{Back:2002wb}.}
\label{Fig.1}
\end{figure}

With these parameters, we calculate the centrality-dependent
multiplicities of charged particles at $\sqrt{s_{NN}}=130, 200$ GeV.
The results are shown in Fig. \ref{Fig.1} and agree with data very
well.

%As a test of our QCM, in Ref. \cite{Shao:2004cn,Yao:2006fk},
%we have well described the transverse momentum
%spectra for the pion, kaon and proton ,
%and the elliptis flows of various of hadrons,
%as the same as the recombination or coalescence models
%\cite{Hwa:2002tu, Hwa:2004ng,Greco:2003xt,Greco:2004rm,Greco:2003mm,
%Fries:2003kq,Fries:2003vb,Nonaka:2003hx}.
%This is what we get by fitting RHIC data for quark or
%antiquark $p_T$ distribution:
%$f(p_T)=(p_T^{2.3}+p_T^{0.2}+1)^{-3.0}$.

In order to compute the distribution of rapidity/pseudorapidity densities
with both energy and centrality dependence,
we have to know the rapidity distribution of quarks and antiquarks
before hadronization.

In the initial state, colliding nuclei are highly Lorentz contracted
along the beam direction. After the initial compression phase, the
evolution of highly excited, and possibly deconfined, strongly
interacting quark matter can be described by the ideal relativistic
hydrodynamics. Under the assumption of full stopping and isentropic
expansion, the amount of entropy ($dS$) contained within the (fluid)
rapidity element $dy$ in the Landau hydrodynamic picture is given by
\cite{Landau:1953gs,Belenkij:1956cd,Mohanty:2003va,
Srivastava:1992gh,Srivastava:1992cg},
\begin{equation}
\frac{dS}{dy} = - \pi R^{2}l s_{0} \beta c_{s} \exp (\beta \omega_{f})
\left[ I_{0}(q) - \frac{\beta \omega_{f}}{q}I_{1}(q) \right] ,
\label{eq18}
\end{equation}
where $2\beta \equiv (1-c_{s}^{2})/c_{s}^{2}$ and $q \equiv
\sqrt{\omega_f^{2} - c_{s}^{2}y^{2}}$ with $c^2_s=(\frac{\partial
P}{\partial \epsilon})_{\rm isentropic}$ the sound velocity square.
$\omega_{f}$ is related to the initial and freeze-out temperature
$T_f$ and $T_0$ by $\omega_{f} \equiv \ln (T_f/T_0)$. $R$ is the
radius of the nuclei. $2l$ is the initial longitudinal length.
$s_{0}$ is the initial entropy density. $I_{0}$ and $I_{1}$ are the
Bessel functions. The quantity $\pi R^2 l s_0$ is fixed to normalize
the experimental data at mid rapidity. For $|\omega_f| \gg c_s y$
the quantity $dS/dy$ can be approximated by a Gaussian distribution,
\begin{equation}
\frac{dS}{dy}\,\sim\, \frac{\exp (-\frac{y^2}{2\sigma^2})}
{\sqrt{2\pi\sigma^2}},
\label{eq6prime}
\end{equation}

where
\begin{equation}
\sigma^2=\frac{2|\omega_f|}{1-c_s^2}\approx
\frac{2c_s^2}{1-c_s^4}\ln(\frac{s_{NN}}{2m_pm_\pi}).
\label{eq6prime}
\end{equation}
where $m_p$ is proton mass, $m_\pi$ is pion mass, taken $T_f\approx
m_\pi$. There is only one parameter $c_{s}^{2}$ left to be
determined by experiments. $dS/dy$ is proportional to $dN/dy$
\cite{Mohanty:2002ty}. Therefore, the rapidity distribution for
quarks and anti-quark before hadronization can be written as
\begin{equation}
\frac{dN}{dy}\,\sim\,\exp(-\frac{y^2}{2\sigma^2}),
 \label{19}
\end{equation}
We take the sound velocity square $c_s^2=1/4$ for QGP before
hadronization ($c_s^2=1/3$ for ideal gas). We have used the fact
that all quarks and anti-quarks  are within the rapidity range $y\in
[-4.2,4.2]$ at 130 GeV and 200 GeV at all centralities of
collisions.

\begin{figure}
\includegraphics[scale=0.7]{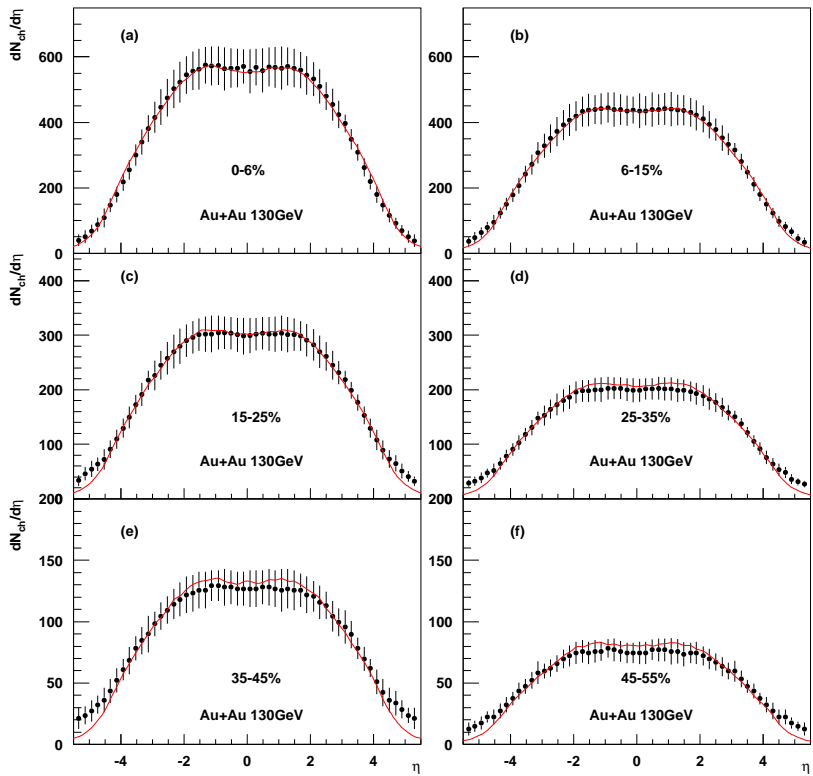}
\caption{Distributions of pseudorapidity density for charged
particles in Au+Au collisions at $\sqrt{s_{NN}}=130$ GeV for six
centrality bins. The solid lines are our results. The data are taken
from PHOBOS \cite{Back:2002wb}.} \label{Fig.2}
\end{figure}

\begin{figure}
\includegraphics[scale=0.7]{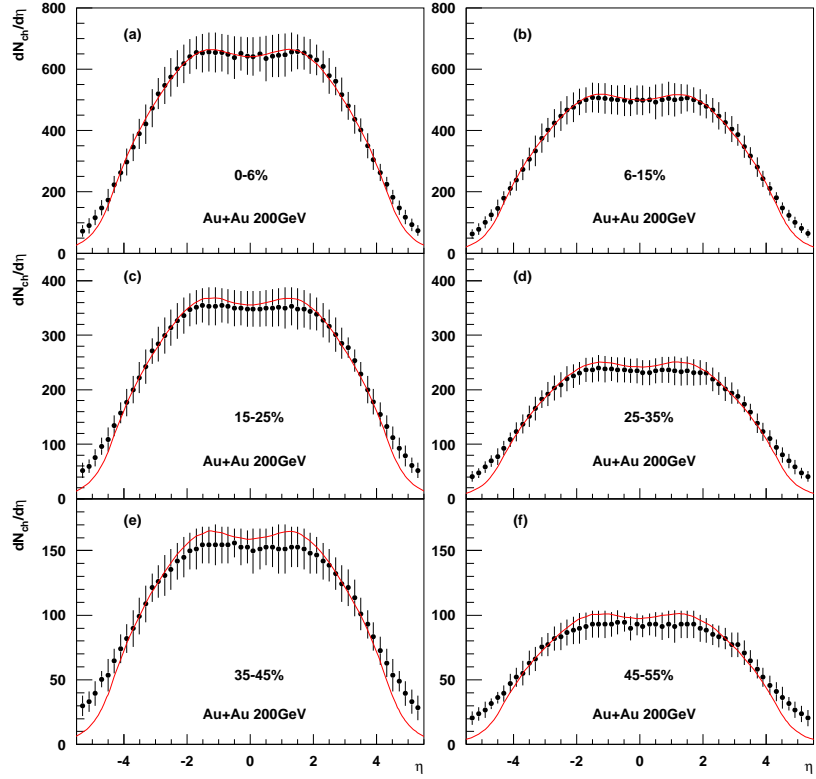}
\caption{Same as Fig.~\ref{Fig.2} but at $\sqrt{s_{NN}}=200$ GeV.}
\label{Fig.3}
\end{figure}

\begin{figure}
\includegraphics[scale=0.5]{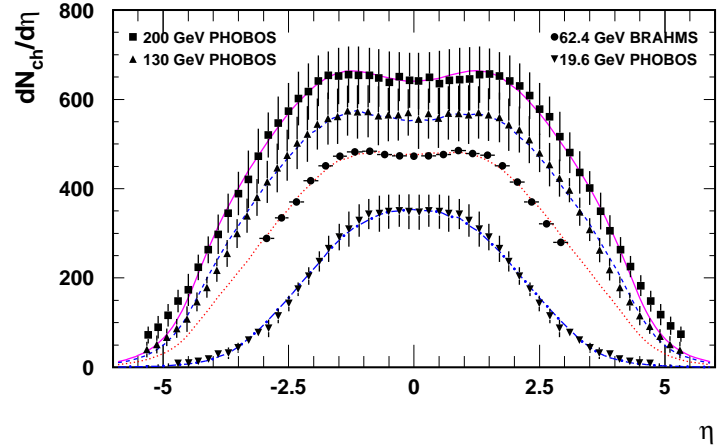}
\caption{Pseudorapidity densities $dN/d\eta$ for charged particles
in most central collisions at various collision energies
$\sqrt{s_{NN}}=19.6, 62.4, 130, 200$ GeV. The lines are our results.
The PHOBOS data are from Ref.~\cite{Back:2002wb}, while
%the STAR data are from Ref.~\cite{Adams:2005cy}, and
the BRAHMS data are from
Ref.~\cite{Staszel:2005}}. \label{f6}
\end{figure}

With this input, we give $dN_{ch}/d\eta$ as functions of $\eta$ at
all available centralities at $\sqrt{s_{NN}}=130, 200$ GeV. The
results are shown in Fig.~\ref{Fig.2} and Fig.~\ref{Fig.3}. One can
see a good agreement between our model predictions and data in
central collisions. For peripheral collisions, there is a slight
deviation from data. The tails at large pseudorapidities especially
in peripheral collisions are associated with remnants of collision
spectators from the incoming nuclei. Therefore our results are
slightly lower than data. Now we study the energy dependence of the
shape of the $\eta$ distribution of charged particles at various
energies of Au +Au collisions. We compute pseudorapidity densities
in full pseudorapidity range in most central collisions at
$\sqrt{s_{NN}}=62.4$ GeV and compare with BRAHMS
data~\cite{Staszel:2005}.
%\cite{Adams:2005cy,Staszel:2005}.
Here, the quark rapidity range is also $y\in [-4.2,4.2]$ and the
sound velocity square is also $c_s^2=1/4$ same as in $130$ and $200$
GeV. By studying we find that the shape of $\eta$ distribution is
mainly determined by the energy and the sound velocity and the the
quark rapidity region only influences the forward pseudorapidity.
The quark rapidity range is possible dependent of collision energy
but not sensitive to that. The calculation results show that there
is same value of the sound velocity for QGP in different collision
energies above. This indicates a certain kind of universality for
the quark matter produced in heavy-ion collisions at the late stage
of evolution (before hadronization) at collision energies from
$62.4$ to $200$ GeV. We also study the results at $19.6$ GeV. We
find that the model predictions disagree with the data using the
constant value of the sound velocity square $c_s^2=1/4$ no matter
how the quark rapidity region is chosen. We have to set the sound
velocity $c_s^2=1/7$ and quark rapidity range $y\in [-3.2,3.2]$. The
change in the sound velocity might reflect the different properties
of the matter produced at $19.6$ GeV and at $62.4$ GeV or higher
energies. The model predictions are shown in Fig.~\ref{f6} at
$\sqrt{s_{NN}}=19.6, 62.4, 130, 200$ GeV. The agreement with data is
also satisfactory, which means that our model captures the energy
behavior in the available collision energies.

\begin{figure}
\includegraphics[scale=0.4]{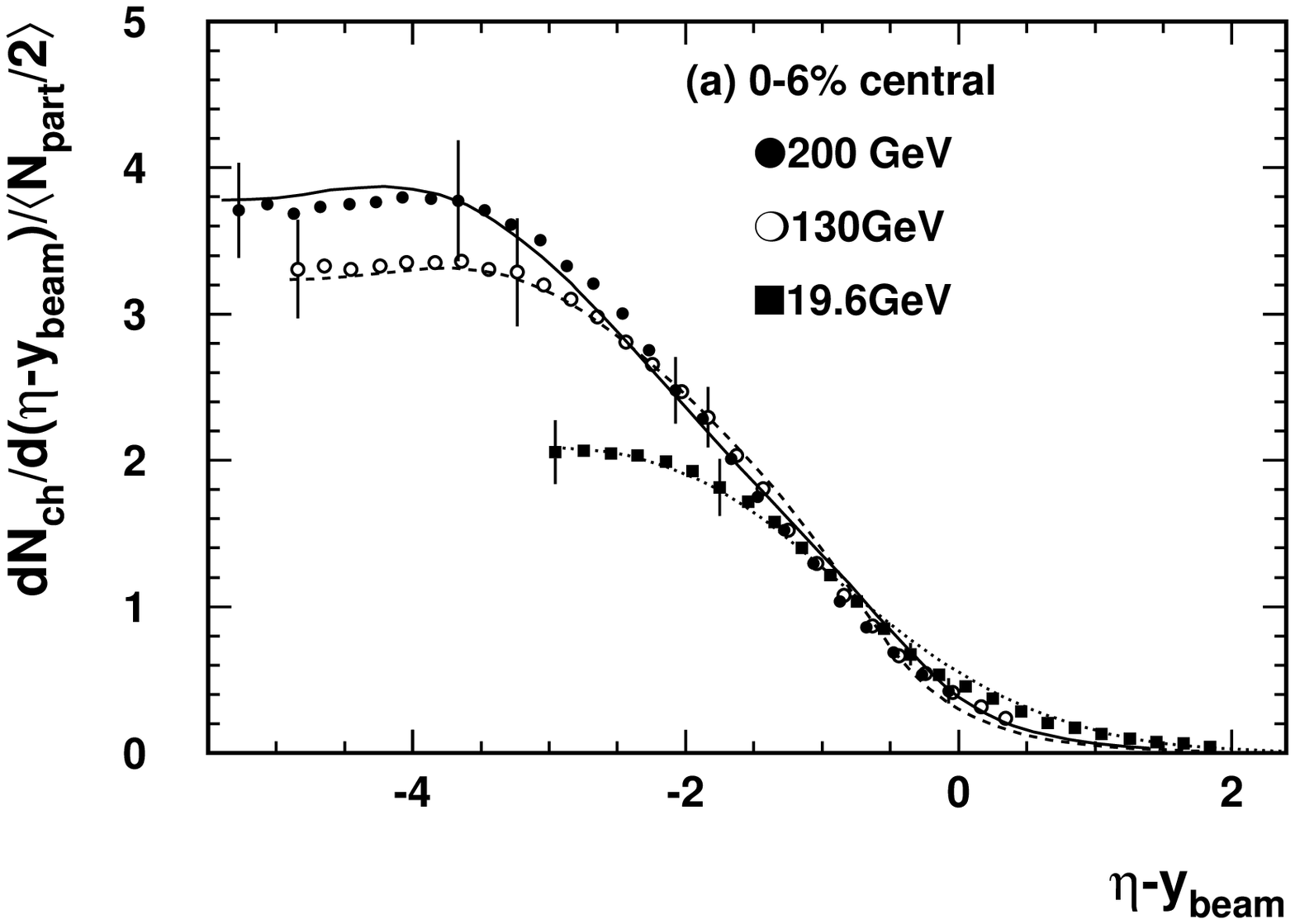}
\includegraphics[scale=0.4]{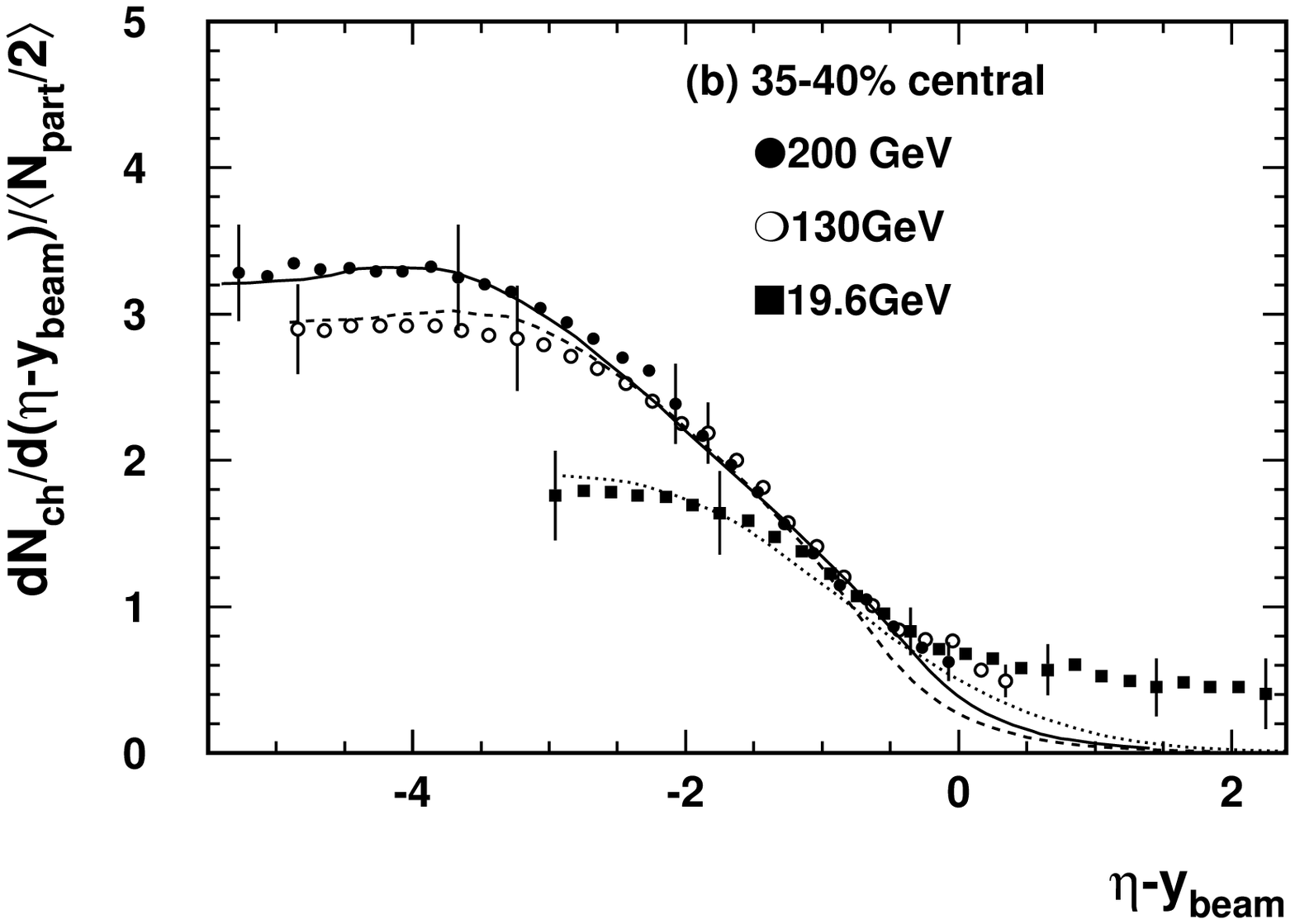}
\caption{The scaled, shifted pseudorapidity rapidity density at
$\sqrt{s_{NN}}=$ 19.6, 130 and 200 GeV. The results at two
centrality bins are presented: 0-6\% and 35-40\%. The lines are our
results. The data are from Ref.~\cite{Back:2002wb},} \label{f7}
\end{figure}

To separate the trivial kinematic broadening of the $dN_{ch}/d\eta$
distribution from the more interesting dynamics, we also study the
scaled, shifted pseudorapidity distribution $dN_{ch}/d\eta'/\langle
N_{part}/2 \rangle$ , where $\eta'=\eta -y_{beam}$ , for Au+Au
collisions at different energies. The calculation results are shown
in Fig.~\ref{f7} at $\sqrt{s_{NN}}=19.6, 130, 200$ GeV and two
centrality bins 0-6\% and 35-40\%. In most central collisions our
model can describe the limit fragmentation very well. But in
peripheral collisions there is a disagreement close to beam
rapidity. The reason is that in peripheral collisions the beam
rapidity region is dominated by spectators and is not covered by our
model.

%\begin{figure}
%\includegraphics[scale=0.4]{eta130-1.eps}
%\includegraphics[scale=0.4]{eta130-2.eps}
%\includegraphics[scale=0.4]{eta130-3.eps}
%\includegraphics[scale=0.4]{eta130-4.eps}
%\includegraphics[scale=0.4]{eta130-5.eps}
%\includegraphics[scale=0.4]{eta130-6.eps}
%\caption{Distributions of pseudorapidity rapidity density for
%charged particles in Au+Au collisions at $\sqrt{s_{NN}}=130$ GeV for six
%centrality bins. The solid lines are our results. The data are taken
%from PHOBOS \cite{Back:2002wb}.}
%\label{Fig.2}
%\end{figure}

%\begin{figure}
%\includegraphics[scale=0.4]{eta200-1.eps}
%\includegraphics[scale=0.4]{eta200-2.eps}
%\includegraphics[scale=0.4]{eta200-3.eps}
%\includegraphics[scale=0.4]{eta200-4.eps}
%\includegraphics[scale=0.4]{eta200-5.eps}
%\includegraphics[scale=0.4]{eta200-6.eps}
%\caption{Same as Fig.~\ref{Fig.2} but at $\sqrt{s_{NN}}=200$ GeV.}
%\label{Fig.3}
%\end{figure}

%\begin{figure}
%\includegraphics[scale=0.4]{y200.eps}
%\includegraphics[scale=0.4]{y130.eps}
%\caption{Distributions of rapidity density for charged particles
%in Au+Au collisions at $\sqrt{s_{NN}}=130, 200$ GeV for six
%centrality bins from $0-6\%$ (top) to $45-55\%$ (bottom).
%The results are our model predictions.}
%\label{Fig.4}
%\end{figure}

%\section{rapidity densities $dN/dy$ and yields of charged pions and kaons}

\begin{figure}
\includegraphics[scale=0.4]{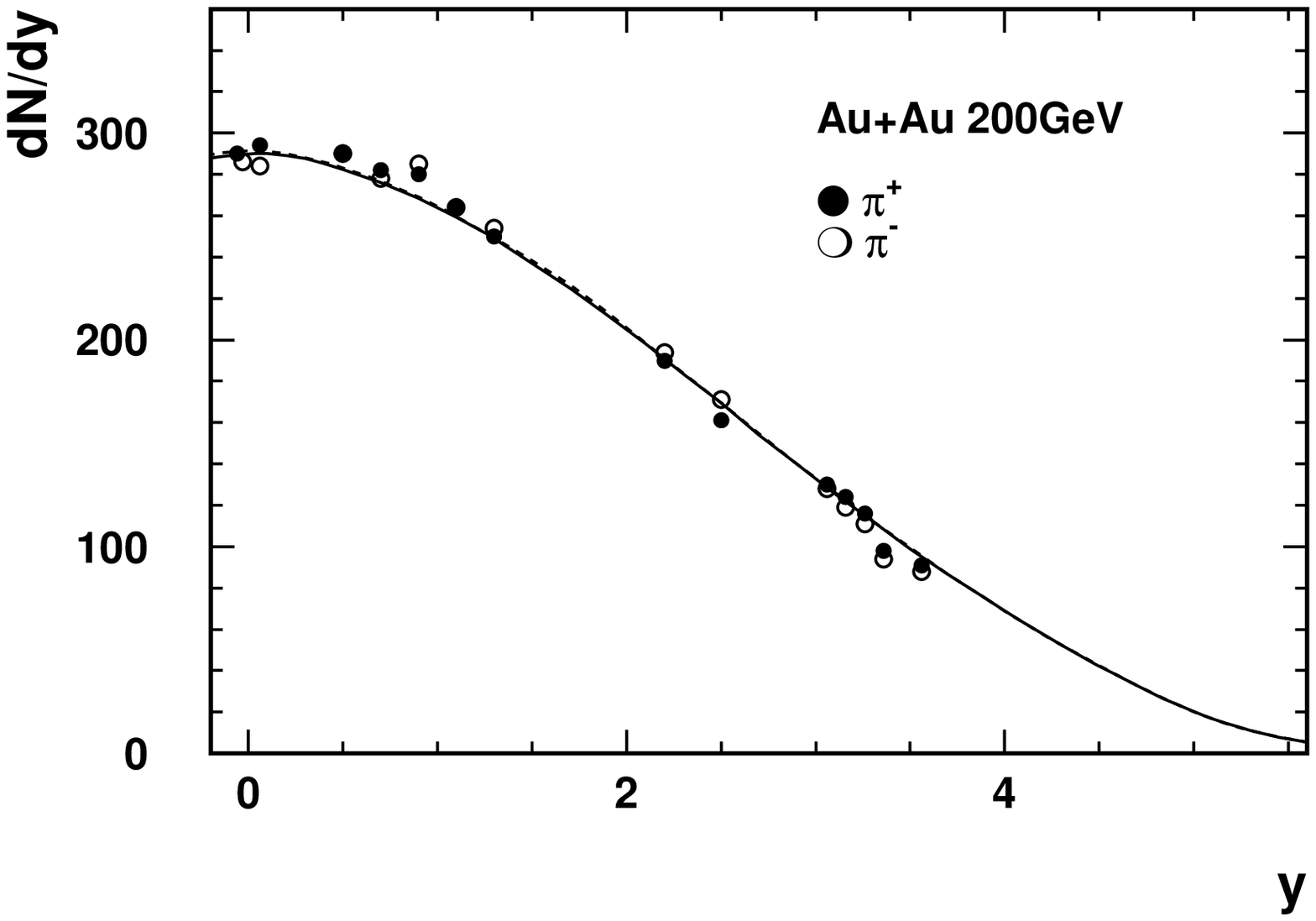}
\includegraphics[scale=0.4]{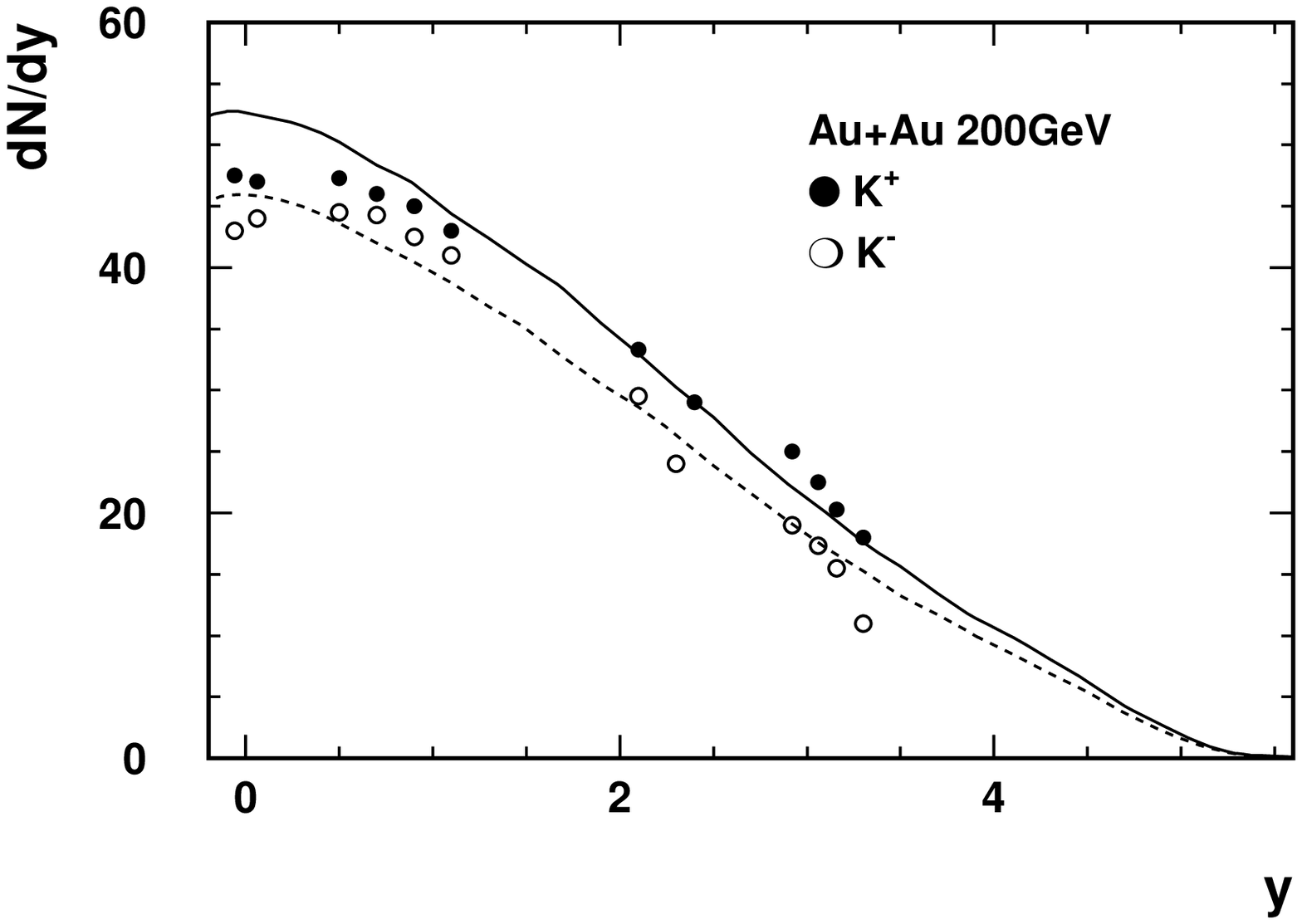}
\caption{ $dN/dy$ in the $0-5\%$ most central collisions at
$\sqrt{s_{NN}}=200$ GeV. The solid lines are $\pi^+$ and $K^{+}$,
and dashed lines are $\pi^-$ and $K^{-}$. The data are given by
BRAHMS collaborations \cite{Bearden:2004yx}. } \label{f8}
\end{figure}

In ultra-relativistic heavy ion collisions at RHIC energies, charged
pions and kaons are copiously produced. The yields of these light
mesons carry the information on the entropy and strangeness created
in the reactions. Here, We calculate the rapidity density $dN/dy$
and yields of charged pions and kaons in full rapidity for central
Au+Au collisions ($0-5\%$) at $\sqrt{s_{NN}}=200$ GeV. The yields of
charged pions and kaons compared with BRAHMS data
\cite{Bearden:2004yx} are shown in Tab. I.  The results for rapidity
density distributions of charged pions and kaons are shown in
Fig.~\ref{f8}.  Here, the pion yields are collected excluding the
contributions of hyperon ($\Lambda$) and kaon $K_{0s}$ decays. One
can see that our model can well describe rapidity densities $dN/dy$
and yields of charged pions and kaons in the whole rapidity range
for central Au+Au collisions at $\sqrt{s_{NN}}=200$ GeV.

\section{Summary and discussions}
\label{summ}

We study in a combination model the rapidity and pseudorapidity
densities at various collision energies and centralities. We use the
Landau relativistic hydrodynamic model to describe the the evolution
of highly excited and possibly deconfined quark matter. As a result,
we obtain the Gaussian-type rapidity spectra of constituent quarks
before hadronization. Then we use our combination model to describe
the hadronization of initially produced hadrons including
resonances, whose decays are dealt with by the event generator
PYTHIA 6.3 \cite{Sjostrand}.
%The main input
%is the Gaussian-type rapidity spectra of constituent quarks as a
%result of Landau hydrodynamic evolution of the fireball.
We compute charged multiplicities and pseudorapidity densities at a
variety of centralities at $130$ and $200$ GeV. The results for
pseudorapidity densities are in good agreement with data in central
collisions. In peripheral collisions, our predictions are slightly
lower than data due to the fact that our model does not include the
influence of the spectators. Our model can well describe the
dependence of pseudorapidity densities and charged multiplicities on
centralities and the number of participants respectively. We also
calculate pseudorapidity densities at $19.6$ and $62.4$ GeV which
describes the RHIC data very well. This means that our model can
reproduce the collision energy dependence of pseudorapidity
densities. However, We find that the value of the sound velocity
square $c_s^2=1/7$ at $19.6$ GeV is different from that at $62.4$
GeV or higher energies. This imply that there is large change in
properties of the hot and dense matter produced at collision energy
between $19.6$ and $62.4$ GeV. To separate the trivial kinematic
broadening of the distributions of the pseudorapidity density from
more interesting dynamics, we compute the scaled and shifted
pseudorapidity density distributions $dN_{ch}/d\eta '/(N_{part}/2)$
with $\eta '=\eta -y _{beam}$ at collision energies 19.6, 130 and
$200$ GeV. The good agreement with data is found except in the beam
rapidity range of peripheral collisions, where our predictions are
lower than data. Finally we present our results for rapidity
densities of charged pions and kaons in most central collisions at $
200 $ GeV. No contradiction to data is found. Note that the BRAHMS
pion data do not include the decay products of $K_s^0$ and
$\Lambda$, we also make the same corrections.

\subsection*{Acknowledgments}

The authors thank Q. Wang, S.-Y. Li, and Z.-T. Liang for helpful
discussions. The work is supported in part by the National Natural
Science Foundation of China under the grant 10475049, the foundation
of University Doctorate Educational Base of Ministry of Education
under the grant 20030422064, and the science fund of Qufu Normal
University.

\end{document}